\documentstyle[12pt,amssymb,amsmath,fleqn]{article}
\begin{document}

{\Large

\noindent {\bf Quasi-Fibonacci oscillators}}

\bigskip

\noindent{\bf A.M. Gavrilik, I.I. Kachurik, A.P. Rebesh}

\noindent Bogolyubov Institute for Theoretical Physics, Kiev 03680,
Ukraine

\noindent E-mail: omgavr@bitp.kiev.ua

\medskip

\begin{abstract}
 We study the properties of sequences of the energy eigenvalues for
some generalizations of $q$-deformed oscillators including the
$p,\!q$-oscillator, the 3-, 4- and 5-parameter deformed oscillators
given in the literature. It is shown that most of the considered
models belong to the class of so-called Fibonacci oscillators for
which any three consecutive energy levels satisfy the relation
$E_{n+1}=\lambda E_n+\rho E_{n-1}$ with real constants $\lambda
,\rho$.
 On the other hand, for certain $\mu$-oscillator known from 1993
 we prove the fact of its non-Fibonacci nature.
  Possible generalizations of the three-term Fibonacci relation are
  discussed among which we choose, as most adequate  for the $\mu$-oscillator,
  the so-called quasi-Fibonacci (or local Fibonacci) property of the energy levels.
  The property is encoded in the three-term quasi-Fibonacci (QF) relation with non-constant,
  $n$-dependent coefficients $\lambda$ and $\rho$.
  Various aspects of the QF relation are elaborated
  for the $\mu$-oscillator and some of its extensions.
\end{abstract}


\bigskip






\medskip

\section{Introduction}\hspace{5mm}
In 1991, the two-parameter family of $p,q$-deformed quantum
oscillators has been introduced \cite{Cha-Ja}. Soon after, in
\cite{Ar_Fib} the family was named "Fibonacci oscillators", due to
the basic property of the family encoded in the three-term linear
recurrence relation \cite{Ar_Fib}
\begin{equation}
E_{n+1}=\lambda E_n+\rho E_{n-1}, \hspace{12mm} n\geq 1,
\hspace{12mm} \lambda, \rho \ \epsilon  \ \mathbb{R} ,\label{1}
\end{equation}
for any three consecutive values of energy levels from the spectrum
of respective deformed quantum oscillators, with $\lambda$ and
$\rho$ definite coefficients that depend on specific model of
deformed oscillator. The usual Fibonacci numbers form the sequence
generated by equation like (\ref{1}) in which $\lambda=\rho=1$ and
the first two members of the sequence are fixed as 1 and 1; the
property states that each number in the sequence results from the
sum of two preceding ones.

Following \cite{Ar_Fib}, in our paper we call the Fibonacci
oscillators those oscillators whose any three energies $E_{n-1}$,
$E_n$, $E_{n+1}$ satisfy the Fibonacci relation (FR) given by
(\ref{1}).
 They form the Fibonacci class of oscillators.
 There exist a number of known oscillators that are the Fibonacci ones.
 For instance, usual harmonic oscillator certainly is the Fibonacci
oscillator in the sense of eq. (\ref{1}) with $\lambda=2$ and
$\rho=-1$.
 As already mentioned, the family of $p,\!q$-oscillators which
 contains well-known models of one-parameter $q$-oscillators, belongs to the
 Fibonacci class.
  However, the question naturally arises about possible existence of other models of
deformed oscillators which satisfy (\ref{1}), i.e., possess the
Fibonacci property (FP).
 Recently, in conjunction with generalized Heisenberg algebras \cite{SCRM}
possible extensions of the FP were studied, either in the direction
of nonlinearization of the relation, or towards so-called $k$-step
extension (in particular, Tribonacci sequence), see \cite{Schork}.
  Also, in ref. \cite{GR-5} some classes of non-Fibonacci (namely
  so-called $k$-bonacci) oscillators have been explored.

Our goal in this paper is two-fold.
 First, we study from the viewpoint of possessing the FP some
 multi-parameter extended
 families of deformed oscillators, including 3- and 4-parametric ones from
\cite{Chung,Borzov, Mizrahi} as well as their 5-parameter extended
model given in \cite{Burban}.
 We prove that all these multi-parameter deformations belong to
 the class of Fibonacci oscillators. On the other hand, we examine
 certain as yet not well studied deformation called $\mu$-oscillator,
 which appeared in \cite{Jan}, and present the proof that this model
 lives outside the Fibonacci class (that is, it is non-Fibonacci one).
 This conclusion served for us as motivation to explore possible
 (still linear) extensions{\footnote{For comparison see e.g. ref. \cite{Wei}
 for an interesting example of nonlinear three-term relation giving the values
 of (non-equidistant) area spectrum of 5-dim Gauss-Bonnet black hole.}}
of the FR and to find among them most adequate and natural form of
generalization which is suitable for the $\mu$-oscillator.
 In other words, our second goal is to provide a proper generalization
 of the FP that is adequate for both the $\mu$-oscillator and its several
multi-parameter extensions. The employed particular generalization
of the FP can be termed "quasi-Fibonacci"\ property.
Correspondingly, the $\mu$-oscillator \cite{Jan} and its appropriate
extensions belong to the class of quasi-Fibonacci (QF) oscillators.
In some sense, this can also be viewed as the deformation of quantum
oscillator whose energy values form a "locally-Fibonacci"\ sequence
of values.

\section{Deformed oscillators of the Fibonacci class}\hspace{5mm}
In this section we examine the energy spectra of the models of
multi-parameter deformed oscillators from  \cite{Chung,Borzov,
Mizrahi,Burban}, from the viewpoint of possessing the FP.
 But first it will be useful to remind main facts about some well-known
 models of deformed oscillators.
 General approach for treatment of deformed oscillators
 involves the notion \cite{Man,Mel,Bona} of the structure function
 $f(N)$ or $\varphi(N)$ that serves to define the deformation.
 In terms of the structure function $\varphi(N)$ given by the
 equalities
 \begin{equation}
\varphi(N)=a^{\dagger}a\ , \hspace{8mm} \varphi(N+1)=aa^{\dagger},
\label{2}
\end{equation}
the most general commutation relation can be written in the form
\[\varphi(N+1)-F(N)\varphi(N)=G(N)\ , 
\]
with $F(N)$ and $G(N)$ real functions. Every special choice of
$F(N)$ and $G(N)$ provides, through solving the latter equation
\cite{Mel}, the particular explicit form of the structure function
$\varphi(N)$ and thus the corresponding model of deformed
oscillator. With fixed $\varphi(N)$, the commutation relations for
the generating elements $a$, $a^{\dagger}$ and $N$ of the concrete
model of deformed oscillator are given as
\begin{equation}
[N,a^{\dagger}]=a^{\dagger}, \hspace{8mm} [N,a]=-a\ , \label{3}
\end{equation}
\begin{equation}
aa^{\dagger}-a^{\dagger}a=\varphi(N+1)-\varphi(N). \label{4}
\end{equation}
That means we use the form of basic commutation relation such that
$F(N)=1$ and $G(N)=\varphi(N+1)-\varphi(N)$. In this case, fixation
of the structure function $\varphi(N)$ completely specifies the
(deformed oscillator) model.

\subsection{Well-known Fibonacci oscillators}\hspace{5mm}
Let us recall some most popular models of deformed oscillators which
involve one or two deformation parameters and belong to the
Fibonacci class of oscillators.

 $\bullet$ {\it {Arik-Coon (AC) model} } \cite{AC}

\vspace{2mm} \noindent This most early known model is given by the
commutation relations (\ref{3}) and
\begin{equation}
aa^{\dagger}-qa^{\dagger}a=1. \label{5}
\end{equation}
In what follows, we use for all the models the same definition of
the Hamiltonian as for the usual harmonic oscillator:
\begin{equation}
H=\frac{1}{2}(aa^{\dagger}+a^{\dagger}a)\  \label{6}
\end{equation}
where $\hbar \omega=1$ is meant.
 Using the appropriately modified (deformed) version of the Fock
 space   \cite{Mel,Bona}   wherein $H|n\rangle=E_n|n\rangle$, and
 $\varphi_{AC}(N)|n\rangle=\varphi_{AC}(n)|n\rangle$, the energy spectrum of AC model is
\begin{equation}
E_n=\frac{1}{2}\Bigl(\varphi_{AC}(n+1)+\varphi_{AC}(n)\Bigr)=\frac{1}{2}\Bigl([n+1]_{AC}+[n]_{AC}\Bigr)=
\frac{1}{2}\biggl(  \frac{q^{n+1}-1}{q-1}+\frac{q^n-1}{q-1}\biggr)
\label{7}
\end{equation}
where the definition of the $AC$-type $q$-bracket (for $X$ either a
number or an operator) and the structure function is given as
\begin{equation}
[X]_{AC}=\frac{q^X-1}{q-1}       \hspace{8mm}      \Leftrightarrow
\hspace{8mm} \varphi_{AC}(n)=\frac{q^n-1}{q-1}=[n]_{AC} \label{8}
\end{equation}
so that the latter satisfies (\ref{5}).

One can easily check that the AC-oscillator energies (\ref{7})
satisfy the FR (\ref{1}) if  $\lambda=1+q$ and $\rho=-q$.

\vspace{2mm}
 $\bullet$ {\it {Biedenharn-Macfarlane (BM) model} } \cite{Bied,Mcf}

\vspace{2mm} The model is determined by the commutation relations
(\ref{3}) and the relations
\begin{equation}
aa^{\dagger}-qa^{\dagger}a=q^{-N} ,   \hspace{12mm}
 aa^{\dagger}-q^{-1}a^{\dagger}a=q^N. \label{9}
\end{equation}
The corresponding structure function which solves (\ref{9}) is
\begin{equation}
\varphi_{BM}(N)=\frac{q^N-q^{-N}}{q-q^{-1}}\equiv [N]_{BM}\ ,
\label{10}
\end{equation}
and the energy spectrum reads
\begin{equation}
E_n=\frac{1}{2}\Bigl(\varphi_{BM}(n+1)+\varphi_{BM}(n)\Bigr)=
\frac{1}{2}\biggl(
\frac{q^{n+1}-q^{-(n+1)}}{q-q^{-1}}+\frac{q^n-q^{-n}}{q-q^{-1}}\biggr).
\label{11}
\end{equation}

The BM oscillator satisfies the FR (\ref{1}) with $\lambda=q+q^{-1}$
and $\rho=-1$.

\vspace{2mm}
 $\bullet$  {\it {$p,\!q$-deformed oscillator}} \cite{Cha-Ja}

\vspace{2mm} This model is determined by the commutation relations
(\ref{3}) and the relations
\begin{equation}
aa^{\dagger}-q a^{\dagger}a=p^{N} ,   \hspace{12mm}
\hspace{8mm} aa^{\dagger}-p\,a^{\dagger}a=q^N ,   \label{12}
\end{equation}
possessing the  $q\leftrightarrow p$ symmetry.
 The relevant structure function satisfying (\ref{12}) is
\begin{equation}
\varphi_{p,q}(N)=\frac{q^N-p^N}{q-p}\equiv[N]_{p,q}\ ,   \label{13}
\end{equation}
and hence the energy spectrum reads
\begin{equation}
E_n=
\frac{1}{2}\biggl(\frac{q^{n+1}-p^{n+1}}{q-p}+\frac{q^n-p^{n}}{q-p}\biggr).
\label{14}
\end{equation}
 The model of $p,q$-deformed oscillator is popular enough: it
finds a number of interesting applications, see e.g.
\cite{AdG,SIGMA,Crn,Algin,Algin2,Ben,UFZh,Plethora}.
 From it, by setting $p=1$ the AC-model is retrieved.
 At $p=q^{-1}$ we recover the BM-type oscillator.
 Finally, the peculiar case of $p=q$ corresponds to the so-called
Tamm-Dancoff (TD) deformed oscillator \cite{Odaka,Jagan}.
 This model, though not so popular as the former two, nevertheless
possesses some interesting properties.
 More details concerning the TD-type oscillator model including some of its
properties can be found in the paper \cite{GR-1}.
   Note that the TD oscillator arises also as a particular
 one-dimensional case in the $SU(d)$ covariant $d$-dimensional so-called
 $q$-deformed Newton oscillator \cite{AAW}.

 With (\ref{14}), the two-parameter $p,q$-model of deformed oscillators
 satisfies the FR
 (\ref{1}) for $\lambda=q+p$ and $\rho=-qp$.
 Accordingly, the AC-oscillator is the Fibonacci oscillator with
 $\lambda=1+q$ and $\rho=-q$, whereas the energy levels of the
 BM-oscillator satisfy (\ref{1}) if we take $\lambda=q+q^{-1}$ and $\rho=-1$.
 Finally, to confirm that the TD-model of deformed oscillator
 does satisfy (\ref{1}), we put $\lambda=2q$ and $\rho=-q^2$.

\vspace{5mm}
\subsection{A family of multi-parameter deformed oscillators}\hspace{5mm}
 In this subsection we focus on those deformed multi-parameter
oscillators which possess more then two deformation parameters in
their defining relations and thus generalize the $q$-  and
$p,\!q$-deformed oscillators. It is of special interest to examine
the properties of such deformed oscillators in dependence on the
number of parameters.
 Our main goal is to examine whether these models of oscillators satisfy
the FR (\ref{1}) and in case they don't, to search for a way of
appropriately modifying the FR.

As the example of multi-parameter oscillators consider the
$(p,q,\alpha,\beta,l)$-deformed oscillator family formulated in
 \cite{Burban} and given by the defining relations
\[aa^{\dagger}-q^{l}a^{\dagger}a=p^{-\alpha N-\beta} ,
\hspace{8mm} aa^{\dagger}-p^{-l}a^{\dagger}a=q^{\alpha N+\beta}.
\]
Here $p,q,\alpha,\beta,l$ are the parameters of deformation.
Creation and annihilation operators act upon the Fock space basis
state $|n\rangle$ by the formulas
\begin{equation}
aa^{\dagger}|n\rangle=\frac{q^{\alpha n+\beta+l}-p^{-(\alpha
n+\beta+l)}}{q^l-p^{-l}}|n\rangle\ ,
 \hspace{5mm}
a^{\dagger}a|n\rangle=\frac{q^{\alpha n+\beta}-p^{-(\alpha
n+\beta)}}{q^l-p^{-l}}|n\rangle. \label{16}
\end{equation}
With the Hamiltonian as in (\ref{6}) the eigenvalues which form the
energy spectrum are
\begin{equation}
E_n=\frac12\biggl(\frac{q^{\alpha n +\beta +l}-p^{-(\alpha n +\beta
+l)}}{q^l-p^{-l}}+ \frac{q^{\alpha n +\beta }-p^{-(\alpha n
+\beta)}}{q^l-p^{-l}}\biggr). \label{17}
\end{equation}

\noindent
 Unlike the models mentioned above, for this 5-parameter model
the formula for $E_n$ cannot be written as
$E_n=\frac{1}{2}(\varphi_{n+1}+\varphi_n)$, but rather as
$E_n=\frac{1}{2}(\varphi_n+\varphi_{n+\frac{l}{\alpha}})$; only in
the 4-parameter case wherein $l=\alpha$ the presentation
$E_n=\frac{1}{2}(\varphi_{n}+\varphi_{n+1})$ is retrieved.

The models of deformed oscillators examined in Subsection 2.1
obviously follow from the $(p,q,\alpha,\beta,l)$-deformed oscillator
as particular cases .
 Moreover, the 5-parameter family reduces at $l=1$  to the 4-parameter
 deformed oscillator treated in Ref. \cite{Borzov,Mizrahi} which in its turn
goes over into the three-parameter deformed quantum oscillators
first formulated in  \cite{Chung}.

\vspace{2mm}
\subsection{Fibonacci property of multi-parameter
oscillator models}\hspace{5mm}
 The constants $\lambda$ and
$\rho$ in (\ref{1}) for the AC, BM and $p,q$-models of deformed
oscillators given in Subsection 2.1 can be deduced easily by solving
the corresponding systems of equations. The principal feature of
those models is that the coefficients $\lambda$, $\rho$ are indeed
constant, i.e., they do not depend on the eigenvalue $n$ of the
(particle number) operator $N$.
   Suppose that (\ref{1}) is true for the extended 3-, 4- and
   5-parameter oscillators.   Then the following statement is true.

{\bf Proposition 1.} {\it The five-parameter deformed oscillator
whose energies $E_n$, $n\geq 0$, are given by (\ref{17}), satisfies,
with certain constants $\lambda$, $\rho$ each of the following two
relations}
\begin{equation}
\begin{cases} E_{n+1}=\lambda E_n+\rho E_{n-1}\ , \cr
E_{n+2}=\lambda E_{n+1}+\rho E_{n}\ ,  {\label{18}}
\end{cases}
\end{equation}
{\it and thus belongs to the class of Fibonacci oscillators}.

The proof proceeds through solving for $\lambda$ and $\rho$ the
system (\ref{18}) with account of (\ref{17}). Recall, these
$\lambda$ and $\rho$ should not depend on $n$. As follows from
(\ref{18}),
\begin{equation}
\lambda=\frac{E_{n+2}E_{n-1}-E_{n+1}E_n}{E_{n+1}E_{n-1}-E^2_n}\
,\hspace{5mm}
\rho=\frac{E^2_{n+1}-E_{n+2}E_n}{E_{n+1}E_{n-1}-E^2_n}\ . \label{19}
\end{equation}
Inserting from (\ref{17}) the expression for $E_j$ , $j=n-1, n, n+1,
n+2$, we obtain\
$ \lambda=B/C$\ where
   $B=B(n;p,q,\alpha,\beta,l),\ C=C(n;p,q,\alpha,\beta,l)$, and
\[
B=\left(\frac{q^{\alpha(n+2)+\beta+l}-p^{-\alpha(n+2)-\beta-l} +
q^{\alpha(n+2)+\beta}-
p^{-\alpha(n+2)-\beta}}{q^l-p^{-l}}\right)\times
\]
\[\hspace{5mm} \times\left(\frac{q^{\alpha(n-1)+\beta+l}-
p^{-\alpha(n-1)-\beta-l}+
q^{\alpha(n-1)+\beta}-p^{-\alpha(n-1)-\beta}}{q^l-p^{-l}}\right)-
\]
\[  \hspace{5mm}
-\left(\frac{q^{\alpha(n+1)+\beta+l}-
p^{-\alpha(n+1)-\beta-l}+q^{\alpha(n+1)+\beta}-
p^{-\alpha(n+1)-\beta}}{q^l-p^{-l}}\right)\times
\]
\[
\hspace{5mm} \times\left(\frac{q^{\alpha n+\beta+l}-p^{-\alpha
n-\beta-l}+q^{\alpha n+\beta}-p^{-\alpha
n-\beta}}{q^l-p^{-l}}\right)\ ,
\]
\[
C=\left(\frac{q^{\alpha(n+1)+\beta+l}-p^{-\alpha(n+1)-\beta-l}+q^{\alpha(n+1)+\beta}-
p^{-\alpha(n+1)-\beta}}{q^l-p^{-l}}\right)\times
\]
\[ \hspace{5mm}
\times\left(\frac{q^{\alpha(n-1)+\beta+l}-p^{-\alpha(n-1)-\beta-l}+
q^{\alpha(n-1)+\beta}-p^{-\alpha(n-1)-\beta}}{q^l-p^{-l}}\right)-
\]
\[
\hspace{6mm} -\left(\frac{q^{\alpha n+\beta+l}- p^{-\alpha
n-\beta-l}+q^{\alpha n+\beta}-p^{-\alpha
n-\beta}}{q^l-p^{-l}}\right)^2.
\]
Calculation shows remarkable cancelation of $n$-dependence: we get
$B/C=B'/C'$ where
\[
B'=-q^{3\alpha}p^{2\alpha+l}-q^{3\alpha}p^{2\alpha}-q^{3\alpha+l}p^{2\alpha+l}-
q^{3\alpha+l}p^{2\alpha}-p^{-\alpha}-p^{-\alpha+l}-q^lp^{-\alpha}-
\]
\[
\hspace{7mm}  -q^lp^{-\alpha+l}+q^\alpha
p^l+q^\alpha+q^{\alpha+l}p^l+q^{\alpha+l}+
q^{2\alpha}p^{\alpha+l}+q^{2\alpha}p^\alpha+
q^{2\alpha+l}p^{\alpha+l}+q^{2\alpha+l}p^\alpha\ ,
\]
\[
C'=-p^l-1-q^lp^l-q^l+2q^\alpha p^{\alpha+l}+ 2q^\alpha
p^\alpha+2q^{\alpha+l}p^{\alpha+l}+
\]
\[\hspace{7mm}
\hspace{-1mm}+2q^{\alpha+l}p^\alpha-q^{2\alpha}p^{2\alpha+l}-
q^{2\alpha}p^{2\alpha}-q^{2\alpha+l}p^{2\alpha+l}-
q^{2\alpha+l}p^{2\alpha}\ .
\]
Moreover, it is easy to check that $(q^\alpha+p^{-\alpha})C'=B'$.
Hence finally
\begin{equation}
\lambda=q^\alpha+p^{-\alpha}\ . \label{20}
\end{equation}
The same procedure applied to $\rho$ yields  $\rho=D/E$ where
\[
D=q^{3\alpha}p^\alpha+q^{3\alpha+l}p^\alpha+ q^\alpha
p^{-\alpha}+q^\alpha p^{-\alpha+l}+
q^{\alpha+l}p^{-\alpha}+q^{\alpha+l}p^{-\alpha+l}-\]
\[
\hspace{5mm}
-2q^{2\alpha}p^l-2q^{2\alpha}-2q^{2\alpha+l}p^l-2q^{2\alpha+l}+
q^{3\alpha}p^{\alpha+l}+q^{3\alpha+l}p^{\alpha+l}\ ,
\]
\[
E=-p^l-1-q^lp^l-q^l+2q^\alpha p^{\alpha+l}+ 2q^\alpha
p^\alpha+2q^{\alpha+l}p^{\alpha+l}+
\]
\[
\hspace{6mm} +2q^{\alpha+l}p^\alpha-q^{2\alpha}p^{2\alpha+l}-
q^{2\alpha}p^{2\alpha}-q^{2\alpha+l}p^{2\alpha+l}-q^{2\alpha+l}p^{2\alpha}\
,
\]
and it is easily checked that $D=(-q^\alpha p^{-\alpha}) E$. As
result,
\begin{equation}
\rho=-q^\alpha p^{-\alpha}\ . \label{21}
\end{equation}
So we conclude: thus found $\lambda$ and $\rho$ in (\ref{20}) and
(\ref{21}) do solve (\ref{18}), and do not depend on $n$.
 Hence we have proven the Proposition.

\section{A case for generalizing the Fibonacci property}\hspace{5mm}
 In this section we will examine, from the
viewpoint of (non)validity of FP, the $\mu$-deformed oscillator
formulated in \cite{Jan}. We will prove that the FP is not valid in
that case. Then our next goal is to find generalization of the FP
appropriate for the $\mu$-oscillator.
 For proper extension, we adopt the so-called "quasi-Fibonacci"\ (QF)
property encapsulated in the QF relation satisfied by members of the
QF sequence. But, first let us recall necessary setup of the
$\mu$-deformed oscillator.

\subsection{Nonlinear $\mu$-deformed oscillator} \hspace{5mm}
The $\mu$-oscillator \cite{Jan} is defined in terms of the unital
algebra whose generating elements $a$, $a^{\dagger}$ and $N$
satisfy
\[
[N,a]=-a\ , \hspace{10mm} [N,a^{\dagger}]=a^{\dagger} \ ,
\hspace{10mm} [N,aa^{\dagger}]=[N,a^{\dagger}a]=0\ ,
\]
\[aa^{\dagger}-a^{\dagger}a=\varphi_{\mu}(N+1)-\varphi_{\mu}(N)\ ,\]
and its structure function is given as
\begin{equation}
a^{\dagger}a=\varphi_{\mu}(N)\equiv\frac{N}{1+\mu N}, \hspace{10mm}
aa^{\dagger}=\varphi_{\mu}(N+1)\equiv\frac{N+1}{1+\mu(N+1)}
\label{22}
\end{equation}
with $\mu$ a deformation parameter.
  That is, the basic commutation relations for the $\mu$-oscillator
  are eq. (\ref{3}) and
\begin{equation}
[a,a^{\dagger}]=\frac{N+1}{1+\mu (N+1)}-\frac{N}{1+\mu N}\ .
\label{23}
\end{equation}
In the $\mu$-deformed version of Fock space, with normalized ground
state $|0\rangle$ such that
\begin{equation}
a|0\rangle=0, \hspace{5mm} N|0\rangle=0, \hspace{5mm}
\varphi_{\mu}(0)=0, \label{24}
\end{equation}
we have the infinite set of basis states
\begin{equation}
|n\rangle=\frac{(a^{\dagger})^n}{\sqrt{\varphi_{\mu}(n)!}}|0\rangle\
, \hspace{8mm} \langle n|m \rangle=\delta_{mn}, \hspace{8mm}
n,m=0,1,2,..., \label{25}
\end{equation}
$\varphi_{\mu}(n)!=\varphi_{\mu}(n)\varphi_{\mu}(n-1)...\varphi_{\mu}(1)$
on which
\[ N|n\rangle=n|n\rangle, \hspace{8mm} \varphi_{\mu}(N)|n\rangle=\varphi_{\mu}(n)|n\rangle\ , \]
\[\langle n-1|a|n \rangle=\langle
n|a^{\dagger}|n-1\rangle=\sqrt{\varphi_{\mu}(n)}=\Bigl(\frac{n}{1+\mu
n}\Bigr)^{\frac{1}{2}}.\]
 Since
\begin{equation}
\varphi_{\mu}(n)=\frac{n}{1+\mu n}\ , \label{26}
\end{equation}
the energy spectrum of this model reads
\begin{equation}
E_n=\frac12\Bigl(\varphi_{\mu}(n+1)+\varphi_{\mu}(n)\Bigr)=\frac{1}{2}\biggl(
\frac{n+1}{1+\mu(n+1)}+\frac{n}{1+\mu n}\biggr) \label{27}
\end{equation}
where the deformation parameter $\mu$ is assumed to satisfy $\mu\geq
0$. Setting $\mu=0$ recovers the known formulae for the standard
quantum harmonic oscillator.

\subsection{Non-Fibonacci nature of the $\mu$-oscillator}
\hspace{5mm}
Suppose the $\mu$-oscillator, with $\varphi(N)$ given in (\ref{22}),
obeys the FP stating that
\begin{equation}
E_{n+1}-\lambda E_n-\rho E_{n-1}=0 \label{28}
\end{equation}
with {\it constant} $\lambda$ and $\rho$. With account of (\ref{27})
this rewrites as
\begin{equation}
\frac{n+2}{1+\mu(n+2)}+\frac{n+1}{1+\mu(n+1)}- \lambda \biggl(
\frac{n+1}{1+\mu(n+1)}+\frac{n}{1+\mu n}\biggr)- \rho
\biggl(\frac{n}{1+\mu n}+\frac{n-1}{1+\mu (n-1)}\biggr)=0\ .
\label{29}
\end{equation}
Then the following statement is true.

\vspace{2mm}
  {\bf Proposition 2.} {\it The $\mu$-oscillator is not the
Fibonacci one, i.e., (\ref{29}) fails for it.}

\vspace{1mm}
 To prove, we take into account the equality, stemming from (\ref{29}),
 of the two polynomials one of which is zero.
 This leads to the following set of equations:

\vspace{3mm} $n^4$:\ \ \ \ $-1+\lambda+\rho=0$\ ;

\vspace{3mm} $n^3$:\ \ \ \ $\lambda(3+2\mu)+\rho(3+2\mu)-3-2\mu=0$\
;

\vspace{3mm} $n^2$:\ \ \ \
$\lambda(6+9\mu-2\mu^2)+\rho(6+7\mu-2\mu^2)- 6-11\mu+2\mu^2=0$\ ;

\vspace{3mm} $n^1$:\ \ \ \
$\lambda(\mu^{2}(\mu^{-1}-1)+(\mu^{-1}+2)(2\mu^{-2}+\mu^{-1}-2))+
\rho(-\mu^{2}(\mu^{-1}+1)+$

\vspace{3mm}\hspace{6mm} \ \ \
$+(\mu^{-1}+2)(2\mu^{-2}-\mu^{-1}-2))-2-10\mu+\mu^{2}+4\mu^3=0$\ ;

\vspace{3mm} $n^0$:\ \ \ \
$2\mu^2-2+\mu(1+2\mu)[(\lambda-1)(\mu^{-1}-1)+\rho(\mu^{-1}+1)]=0$\
. \vspace{2mm}

\noindent We have to solve this set for $\lambda$ and $\rho$.
   The first two equations yield $\lambda+\rho=1$.
However, insertion this in the rest of equations leads us to
inconsistency, as there is no solution satisfying the whole system,
for arbitrary $\mu$. So we conclude: the $\mu$-oscillator with
$\varphi(n)$ and $E_n$ from (\ref{26})-(\ref{27}) {\em does not
possess the Fibonacci property} (\ref{28}), with constant $\lambda$
and $\rho$.

Therefore, we have to find a modification of the Fibbonacci property
which would be adequate for the $\mu$-oscillator and also for some
models that extend it.
\begin{center}
\it{Possible ways of generalizing the FP}
\end{center}
Among possible modifications, we could try an extended version of
the usual Fibonacci relation (\ref{1}) obtained merely by adding
extra terms. Say we could consider the $k$-term extended (so-called
$k$-bonacci) relation of the form (see e.g. \cite{Schork,GR-5}):
\[
E_{n}=\alpha_1 E_{n-1}+\alpha_2E_{n-2} +\alpha_3
E_{n-3}+...+\alpha_{k} E_{n-k}\ .
\]
For the $\mu$-oscillator, let us check the first extended case of
$k=3$ (so-called "Tribonacci relation"). It is proved that the
$\mu$-oscillator does not satisfy the Tribonacci relation. The proof
goes in complete analogy with the above case of $k=2$, see
Proposition 2: by deriving the corresponding system of (now six)
equations and then trying to solve that system. As result, we are
led to the inconsistency of the system and thus to the conclusion
that the $\mu$-oscillator does not satisfy the Tribonacci relation.
Moreover, this negative result for the $\mu$-oscillator extends to
all other cases $k\geq 3$ of $k$-bonacci relation.

Next, the most radical possibility would be to search for some
nonlinear generalization of (\ref{1}), say in the form
$E_{n+1}=F(E_n, E_{n-1})$ with $F(x, y)$ an appropriate function
\cite{SCRM}. We prefer however to {\it preserve both the linearity
and the three-term form} of relation. But some price should be payed
for such a choice, and the price is nothing but the loss of constant
nature of $\lambda$ and $\rho$. In other words, these coefficients
inevitably should be $n$-dependent.

\subsection{Quasi-Fibonacci property of $\mu$-deformed oscillator}\hspace{5mm}
 We will call "quasi-Fibonacci"\ oscillators those deformed
oscillators the energy spectrum of which satisfy the extended or
quasi-Fibonacci (QF) relation involving
$\lambda=\lambda(n)\equiv\lambda_n$ and $\rho=\rho(n)\equiv\rho_n$:
\begin{equation}
E_{n+1}=\lambda_n E_n+\rho_n E_{n-1}. \label{30}
\end{equation}
We will show that the $\mu$-deformed oscillator belongs to the set
of QF oscillator models. That is, by passing from the Fibonacci
property to the QF one, we will prove that it is the QF property
which is adequate to the energy spectrum of $\mu$-oscillator.

There exist three different ways of finding $\lambda_n$ and $\rho_n$
needed to prove the QF property.

\subsubsection{ First way of finding
         $\lambda_n$, $\rho_n$ (by the "splitting" ansatz)}\hspace{5mm}
To start with, let us recall that the energy spectrum $E_n$ of the
deformed oscillator consists of two terms given by the structure
function $\varphi(n)$:
\begin{equation}
E_n=\frac{1}{2}\bigl(\varphi(n)+\varphi(n+1)\bigr)\ . \label{31}
\end{equation}
From (\ref{30}) and (\ref{31}), using the "splitting", we have the
following system of equations:
\begin{equation}
\begin{cases} \varphi(n+1)=\lambda_n \varphi(n)+\rho_n \varphi(n-1) , \cr
\varphi(n+2)=\lambda_n \varphi(n+1)+\rho_n \varphi(n) . \label{32}
\end{cases}
\end{equation}
Solving the latter system, for $\lambda_n$ and $\rho_n$ we have
\begin{equation}
\lambda_n=\frac{\varphi(n+1)-\rho_n\,\varphi(n-1)}{\varphi(n)}\ ,
\hspace{5mm}
\rho_n=\frac{\varphi(n+2)\varphi(n)-\varphi^2(n+1)}{\varphi^2(n)-
\varphi(n+1)\varphi(n-1)}\ . \label{33}
\end{equation}
From (\ref{33}), using the expression (\ref{26}) for $\varphi(n)$ of
the $\mu$-oscillator, we finally obtain
\begin{equation}
\lambda_n=2\frac{(1+\mu(1+2n))}{(1+2\mu n)}\cdot\frac{(1+\mu
n)}{1+\mu(n+2)}\ , \label{34}
\end{equation}
\begin{equation}
\rho_n=-\frac{(1+2\mu(n+1))}{(1+2\mu n)}\cdot \frac{(1+\mu
n)(1+\mu(n-1))}{(1+\mu(n+2))(1+\mu(n+1))} \ . \label{-35}
\end{equation}
It is clear that, for consistency, from (\ref{34}) and (\ref{-35})
in the limit $\mu\rightarrow 0$ we should recover the {\it
constants}:
\begin{equation}
\lim_{\mu\rightarrow 0}\lambda_n=2\ , \hspace{5mm}
\lim_{\mu\rightarrow 0}\rho_n=-1\ , \label{-36}
\end{equation}
i.e., the values $\lambda$, $\rho$ of the usual harmonic oscillator.
Obviously, that is true. Hence we conclude that the $\mu$-oscillator
satisfies (\ref{30}) with (\ref{27}) and (\ref{34}), (\ref{-35}),
and thus is quasi-Fibonacci one.

\subsubsection{Second way to find $\lambda_n$ and
$\rho_n$ (by the substitution ansatz)}\hspace{5mm}
 Recall that most general form of defining commutation
relation looks as \cite{Mel}
\begin{equation}
\varphi(N+1)-F(N)\varphi(N)=G(N)\ . \label{-37}
\end{equation}
By treating this as the recursion relation and fixing the initial
conditions
\begin{equation}
\varphi(0)=0, \hspace{5mm} \varphi(1)=G(0)=\varepsilon\
,\hspace{5mm} \varepsilon \ \epsilon \ \Re, \label{38}
\end{equation}
the structure function $\varphi(n)$ can be obtained as
\begin{equation}
\varphi(n)=F(n-1)!\Biggl(\varepsilon+\sum^{n-1}_{j=1}\frac{G(j)}{F(j)!}\Biggr)\
,\label{39}
\end{equation}
where $F(j)!=F(j)F(j-1)...F(1)$ and $F(0)!=1$ (at $\varepsilon=0$
the corresponding formula has been given in \cite{Mel}).

For our aim, we proceed in analogy with (\ref{-37})-(\ref{39}). So
we use the ansatz
\begin{equation}
\rho_n=\lambda_{n-1} \hspace{5mm} {\rm or} \hspace{5mm}
\rho_{n+1}=\lambda_{n}. \label{40}
\end{equation}
With account of this, the QF relation (\ref{30}) rewrites in the
form
\begin{equation}
E_{n+2}=\lambda_{n+1}E_{n+1}+\lambda_nE_n\  \label{41}
\end{equation}
or, for more explicit analogy with (\ref{-37}), in the form
\begin{equation}
\lambda_{n+1}-\Bigl(-\frac{E_n}{E_{n+1}}\Bigr)\lambda_n=\frac{E_{n+2}}{E_{n+1}}\
,
 \hspace{5mm} n\geq 1 \ . \label{42}
\end{equation}
Fixing the initial conditions as
\begin{equation}
\lambda_0=\lambda(0)=0\ , \hspace{5mm}
\lambda_1=\lambda(1)=\frac{E_2}{E_1} \label{43}
\end{equation}
we derive for $\lambda_n$ the expression
\begin{equation}
\lambda_n=\Bigl(-\frac{E_{n-1}}{E_n}\Bigr)!
\sum^{n-1}_{j=0}\frac{E_{j+2}/E_{j+1}}{(-E_j/E_{j+1})!}\ .
\label{44}
\end{equation}
The coefficient $\rho_n$ then follows from (\ref{40}). In another
way, we may use the relation (\ref{42}) to deduce 
$\lambda_n$ recursively. Recall that the lowest values of
$\lambda_n$ besides (\ref{43}), are:
\[
\lambda_2=-1+\frac{E_3}{E_2}\ , \hspace{28mm}
\lambda_3=-1+\frac{E_4+E_2}{E_3}\ ,
\]
\[
\lambda_4=-1+\frac{E_5+E_3-E_2}{E_4}\ , \hspace{8mm}
\lambda_5=-1+\frac{E_6+E_4-E_3+E_2}{E_5}\ ,
\]
and the formula for generic $\lambda_n$ (which is equivalent to
(\ref{44}), is
\begin{equation}
\lambda_n=-1+\frac{E_{n+1}+E_{n-1}+\sum^{n-2}_{j=2}(-1)^{n-k+1}E_j}{E_n}
=  \frac{1}{E_n} \sum^{n+1}_{j=2}(-1)^{n-j+1}E_j\ . \label{45}
\end{equation}
Let us remark that instead of the initial condition
$\lambda(0)=0$ in eq. (\ref{43}), we can set more general one,
namely
\begin{equation}
\lambda_0=\lambda(0)=c\ , \hspace{5mm} c=E_1/E_0. \label{46}
\end{equation}
Then it follows from (\ref{41}) that
\[
\lambda_1=\frac{E_2-cE_0}{E_1}, \hspace{5mm}
\lambda_2=\frac{E_3-E_2+cE_0}{E_2}\ , \hspace{5mm}
\lambda_3=\frac{E_4-E_3+E_2-cE_0}{E_3}\ ,
\]
and by applying induction, we arrive at the desired result
\begin{equation}
\lambda_n=\frac{\sum_{k=2}^{n+1}(-1)^{n-k+1}E_k+(-1)^ncE_0}{E_n}\ .
\label{47}
\end{equation}
Of course, putting $c=0$ reproduces (\ref{45}). Using $c=E_1/E_0$ in
(\ref{46}) the formula (\ref{47}) takes the form
\begin{equation}
\lambda_n=(-1)^{n+1}\frac{1}{{E_n}}\sum_{k=1}^{n+1}(-1)^{k}E_k
\label{48}
\end{equation}
which after substitution of (\ref{27}) yields the desired result:
\begin{equation}
\lambda_n=\left\{\frac{(-1)^n}{1+\mu}+\frac{n+2}{1+\mu(n+2)}\right\}
\frac{(1+\mu n)(1+\mu(n+1))}{n(1+\mu(n+1))+(n+1)(1+\mu n)}\ ,
\label{49}
\end{equation}
\begin{equation}
\rho_n=\left\{\frac{(-1)^{n-1}}{1+\mu}+\frac{n+1}{1+\mu(n+1)}\right\}
\frac{(1+\mu(n-1))(1+\mu n)}{(n-1)(1+\mu n)+n(1+\mu(n-1))}\ .
\label{50}
\end{equation}
The solution (\ref{49})-(\ref{50}) of eq. (\ref{30}) obviously
differs from the solution (\ref{34})-(\ref{-35}) as the ansatz
applied in subsection 3.3.2 is essentially different from the one
used in 3.3.1.

 \vspace{5mm} \subsubsection{Third (most general) approach for
deriving $\lambda_n$ and $\rho_n$}\hspace{5mm}
 We first remark that the formulae derived in the preceding
 two subsections do not constitute general result. The goal of
 this subsection is just to find $\lambda_n$ and $\rho_n$ from the
 QF relation (\ref{30}) within most general approach.
  If we take instead of (\ref{30}) the system
\begin{equation}
\begin{cases} E_{n+1}=\lambda_n E_n+\rho_n E_{n-1} , \cr
E_{n+2}=\lambda_{n+1} E_{n+1}+\rho_{n+1} E_{n} , \cr
\end{cases} \label{51}
\end{equation}
(as 2 equations with 4 unknowns) we cannot calculate
$\lambda$ and $\rho$ as before, see (\ref{18})-(\ref{19}).

To find explicitly $\lambda_n$ and $\rho_n$ satisfying
(\ref{30}) along with (\ref{27}), that is, the relation
\begin{equation}
 \frac{n+2}{1+\mu(n+2)}=(\lambda_n-1) \frac{n+1}{1+\mu(n+1)}
+\left(\lambda_n + \rho_n \right)  \frac{n}{1+\mu n} + \rho_n
\frac{n-1}{1+\mu(n-1)} \ , \label{52}
\end{equation}
we now follow more general scheme:
we replace (\ref{52}) with the pair of equations
\begin{equation}
\frac{2n+2}{1+\mu(n+2)}=\bigl(\lambda_n-1\bigr)\frac{n+1}{1+\mu(n+1)}+
\frac{K(n;\mu)(n+1)}{1+\mu(n+1)}\ ,\label{53}
\end{equation}
\begin{equation}
\frac{-n}{1+\mu(n+2)}=\bigl(\lambda_n+\rho_n\bigr)\frac{n}{1+\mu
n}+\rho_n\frac{n-1}{1+\mu(n-1)}-\frac{K(n;\mu)(n+1)}{1+\mu(n+1)} \ .
\label{54}
\end{equation}
Note that, the unspecified function $K=K(n;\mu)$ introduced here,
due to its arbitrariness, encapsulates the non-uniqueness of
splitting (\ref{52}) into two equations and thus provides
equivalence of (\ref{52}) with the pair (\ref{53})-(\ref{54}).
 Indeed, adding (\ref{53}) and (\ref{54}) yields (\ref{52}).
 Or, isolating the last term (with $K(n;\mu)$) in
(\ref{53}) and the identical term in (\ref{54}), we get the
equivalence: (\ref{52}) $\leftrightarrow$ (\ref{53})\&(\ref{54}).

Now, (\ref{53})-(\ref{54}) are easily solved for $\lambda_n$ and
$\rho_n$ that yields the desired general result
\begin{equation}\hspace{-8mm}
\lambda_n=1-K(n;\mu)+2\frac{1+\mu (n+1)}{1+\mu(n+2)} \ ,
\label{55}
\end{equation}
\begin{equation}
\rho_n=\frac{1+\mu(n\!-\!1)}{1+2(n\!-\!1)(1+\mu n)} \
\!\left\{K(n;\mu)\,\frac{2(n+1)(1+\mu n)\!-\!1}
        {1+\mu (n+1)}-\frac{4n(1+\mu(n+1))}{1+\mu (n+2)} \right\}.
        \label{56}
\end{equation}

{\bf Remark 1}. Various choices of the function $K(n;\mu)$ are of
interest. Say, by properly specified choice $K=K^{(1)}$
(respectively
$K=K^{(2)}$), we can reproduce from
(\ref{55})-(\ref{56}) the expressions (\ref{34}), (\ref{-35}) in
subsection 3.3.1 (respectively the formulae (\ref{49}), (\ref{50})
in subsection 3.3.2). It is easy to verify that the corresponding
choices of $K(n;\mu)$ are:
\begin{equation}
K^{(1)}=\frac{1+3\mu n(1+2\mu)+2\mu(\mu n^2+1)}{(1+2\mu
n)(1+\mu(n+2))}\ , \label{57}
\end{equation}
\[
K^{(2)}=\frac{\mu}{1+\mu(n+2)}+
 \frac{(-1)^{n+1}(1+\mu n)(1+\mu(n+1))} {(1+\mu)\bigl\{n\,(1+\mu(n+1))+(n+1)(1+\mu
n)\bigr\}}+\
\]
\begin{equation}
\hspace{10mm}+\frac{(n+1)\bigl\{ 3\mu n+(5n+1)(1+\mu n)\bigr\}}
{(1+\mu(n+2))\bigl\{n\,(1+\mu(n+1))+(n+1)(1+\mu n)\bigr\}}\ .
\label{58}
\end{equation}
From viewing the quasi-Fibonacci relation as "locally-Fibonacci"
property, see end of Introduction, it follows that the (auxiliary)
function $K(n;\mu)$, introduced in order to account the
non-uniqueness of replacing (\ref{52}) by the pair
(\ref{53})$\&$(\ref{54}) plays the role somewhat resembling gauge
freedom in gauge theories: the quantities $\lambda_n$ and $\rho_n$
depend on $K(n;\mu)$ while the energies $E_n$ do not, see (\ref{27})
and (\ref{51})-(\ref{52}).

 {\bf Remark 2}. As mentioned above, the usual harmonic oscillator
is Fibonacci oscillator for which $\lambda=2$, $\rho=-1$.
  Its structure function and energy spectrum obviously stem from
  (\ref{26}), (\ref{27})
if $\mu\rightarrow 0$.
 However, sending $\mu\rightarrow 0$ in the formulas (\ref{55})-(\ref{56})
 gives
\begin{equation}
\lambda_n=3-K(n;0),   \hspace{18mm}
\rho_n=\frac{K(n;0)(2n+1)-4n}{2n-1} \ , \label{59}
\end{equation}
i.e., there is a residual dependence on $n$ if $K(n;0)\ne 1$.
This looks as a kind of surprise: due to $n$-dependent $\lambda_n$,
$\rho_n$ the usual oscillator {\it can also be treated as
quasi-Fibonacci oscillator} (satisfying (\ref{30})).
 Only if $K(n;0)=1$ is put, we get $\lambda=2$,
 $\rho=-1$, and the usual oscillator becomes genuine Fibonacci oscillator.

\vspace{2mm}

Concerning the last phrase it is worth to add the following.
>From (\ref{57}) and (\ref{58}) at $\mu\rightarrow0$ we have:
$K^{(1)}|_{\mu=0}=1$, but $K^{(2)}|_{\mu=0}\neq1$. This explains the
fact that $\lambda_n$ and $\rho_n$ from (\ref{34})-(\ref{-35}) do
lead at $\mu\rightarrow0$ to the values $\lambda=2$ and $\rho=-1$ of
the ordinary oscillator, while those in (\ref{49})-(\ref{50}) do not
(the reason is the very ansatz $\rho_n=\lambda_{n-1}$ in
(\ref{40})).

{\bf Remark 3.} The extension based on replacing the constant
$\lambda$, $\rho$ with $\lambda_n=\lambda(n)$ and $\rho_n=\rho(n)$
is of principal value for the QF property. However, for the true QF
property to be valid it is in fact enough
 that only one of the coefficients depends on $n$, the other being constant.
 For instance, consider the case: $\lambda_n=\lambda(n)$ and $\rho=-1$
for the considered $\mu$-oscillator. Imposing $\rho=-1$ in
(\ref{56}) yields the specified $K(n;\mu)$-function:
\[ K(n;\mu)_{\rho=-1} = \frac{(1+\mu(n+1))\{2\mu^2n^2(3n+1)+
4\mu
n(3n+2\mu)+\mu(n-2)+6n-1\}}{(1+\mu(n-1))(1+\mu(n+2))\{2(n+1)(1+\mu
n)-1\}}
\]
With this particular $K$-function, we find the $\lambda_n$ related
to $\rho=-1$:
\begin{equation}
\lambda(n)_{\rho=-1}=1-\frac{(1+\mu(n+1))\{10\mu n^2(2+\mu
n)+(1-4\mu)(3\mu n+1)+12n\}}{(1+\mu(n-1))(1+\mu(n+2))\{2(n+1)(1+\mu
n)-1\}}\ . \label{60}
\end{equation}
It is worth to mention that similar form of the QF relation, with
$\rho=-1$ and certain $\lambda=\lambda(n)$,
appears in \cite{d-Negro}, where the members of the QF sequence
therein have the physical sense of (the Fourier transform of) the
oscillation amplitude of dipole with number $n$ in the non-periodic
chain of dipoles.

Likewise, we may impose $\lambda=2$ and then find from
(\ref{55})-(\ref{56}) the relevant $K$-function and the
corresponding $\rho=\rho_n$:
\[\hspace{-8mm} K(n;\mu)_{\lambda=2}= 2\frac{(1+\mu(n+1))}{(1+\mu(n+2))}-1\ ,
\]
\begin{equation}
\rho(n)_{\lambda=2}=-\frac{1+\mu(n-1)}{1+2(n+1)(1+\mu
n)}\cdot\frac{n(2+5\mu)+4\mu
n(n+\mu)+2\mu^2n^2(n+3)-1}{(1+\mu(n+1))(1+\mu(n+2))}\ . \label{61}
\end{equation}

\vspace{3mm} {\bf Remark 4.} Let $K$ in (\ref{55}), (\ref{56}) be
equal to zero. Then $\lambda_n=\frac{P_1}{Q_1}$, where $P_1$ and
$Q_1$ are linear in $n$, while $\rho_n=\frac{P_3}{Q_3}$ where $P_3$
and $Q_3$ are  cubic in $n$.      At $K=1$, however, $\lambda_n$ is
again "linear/linear"\ expression in $n$, but $\rho_n$ is the
"quadratic/quadratic"\ one. Let us compare the structure of the
coefficients $\lambda_n$ and $\rho_n$ which were obtained in
different ways, see subsections 3.3.1-3.3.3. While $\lambda_n$,
$\rho_n$ in (\ref{34}), (\ref{-35}) are of the form:
$\lambda_n=\frac{P_2^{(1)}}{Q_2^{(1)}}$ and
$\rho_n=\frac{\widetilde{P}_3^{(1)}}{\widetilde{Q}_3^{(1)}}$, the
structure seen in (\ref{49}), (\ref{50}) looks as
$\lambda_n=\frac{P_3^{(2)}}{Q_3^{(2)}}$ and
$\rho_n=\frac{\widetilde{P}_3^{(2)}}{\widetilde{Q}_3^{(2)}}$.
Likewise, an appropriate choice of the function $K(n;\mu)$ in the
general expressions (\ref{55}), (\ref{56}), yields another
particular case for $\lambda_n$, $\rho_n$ such that they also both
take the form $\frac{P_3^{(3)}}{Q_3^{(3)}}$. Indeed, the choice
\[
K=\frac{\bigl(1+\mu(n+1)\bigr)\bigl(1+2(n-1)(1+\mu
n)\bigr)}{\bigl(1+\mu(n+2)\bigr)\bigl(-1+2(n+1)(1+\mu n)\bigr)}
\]
again leads to the "cubic/cubic"\ expressions for both $\lambda_n$
and $\rho_n$:
\[
\lambda_n=1+\frac{1+\mu(n+1)}{1+\mu(n+2)}\cdot\frac{3+2(1+\mu
n)(3n+1)}{1-2(n+1)(1+\mu n)},
\]
\[
\rho_n=-\frac{1+\mu(n-1)}{1+2(n-1)(1+\mu
n)}\cdot\frac{1+2n(1+\mu(n+3))}{1+\mu(n+2)}.
\]

\vskip5mm
 To end this section, we find it instructive to tabulate
 $\lambda_n$ and $\rho_n$ from (\ref{55})-(\ref{56}) for low values of $n$, together with
 $E_n$, at fixed $K(n;\mu)=1$, see Table 1.

 \begin{center}
\noindent{\footnotesize{\bf%
Table 1. Explicit form of \boldmath{$\lambda_n$} and
\boldmath{$\rho_n$} at low values of \boldmath{$n$}.
}}
 \end{center}

\vspace{5mm} \vskip1mm

\tabcolsep20.1pt

\noindent\begin{tabular}{c|c|l|l}
\hline%
\hline%
\rule{0pt}{10pt}
$n$  &   $\lambda_n$ &   $\rho_n$  &   $E_n$ \\
\hline%
\hline%
\rule{0pt}{2pt} \vspace{-3mm}
{} & {} & {} & {}\\
 $0$ & \hspace{-2mm}$\lambda_0=\frac{2+2\mu}{1+2\mu}$ & \hspace{-3mm} $\rho_0=\frac{\mu-1}{1+\mu}$ &
\hspace{-2mm} $E_0=\frac{1}{2(1+\mu)}$ \\
\vspace{-4mm}
{} & {} & {} & {}\\
\hline%
\vspace{-3mm}
{} & {} & {} & {}\\
 $1$ & \hspace{-2mm}$\lambda_1= \frac{2+4\mu}{1+3\mu}$ &  \hspace{-2mm}$\rho_1\!=\frac{3+4\mu}{1+2\mu}-\frac{4(1+2\mu)}{1+3\mu}$ &
 \hspace{-2mm} $E_1=\frac{1}{2(1+\mu)}+\frac{1}{1+2\mu}$ \\
\vspace{-4mm}
{} & {} & {} & {}\\
\hline%
\vspace{-3mm}
{} & {} & {} & {} \\
 $2$ & \hspace{-2mm}$\lambda_2 = \frac{2+6\mu}{1+4\mu}$ & \hspace{-3mm}
 $\rho_2\!=\!\frac{1+\mu}{3+4\mu}\!\left(\!\frac{5+12\mu}{1+3\mu}\!-\!\frac{8(1+3\mu)}{1+4\mu}\!\right)$ \hspace{-3mm} &
\hspace{-2mm} $E_2=\frac{1}{1+2\mu}+\frac{3}{2(1+3\mu)}$ \\
\vspace{-4mm}
{} & {} & {} & {}\\
\hline
\vspace{-3mm}
{} & {} & {} & {}\\
 $3$ & \hspace{-2mm}$\lambda_3= \frac{2+8\mu}{1+5\mu} $  & \hspace{-3mm}
 $\rho_3\!=\!\frac{1+2\mu}{5+12\mu}\!\left(\!\frac{7+24\mu}{1+4\mu}\!-\!\frac{12(1+4\mu)}{1+5\mu}\!\right)$ \hspace{-3mm}&
\hspace{-2mm} $E_3=\frac{3}{2(1+3\mu)}+\frac{2}{1+4\mu}$ \\
\vspace{-4mm}
{} & {} & {} & {}\\
\hline%
\vspace{-3mm}
{} & {} & {} & {}\\
 $4$ & \hspace{-2mm}$\lambda_4= \frac{2+10\mu}{1+6\mu} $ & \hspace{-3mm}
 $\rho_4\!=\!\frac{1+3\mu}{7+24\mu}\!\left(\!\frac{9+40\mu}{1+5\mu}\!-\!\frac{16(1+5\mu)}{1+6\mu}\!\right)$ \hspace{-3mm}&
\hspace{-2mm} $E_4=\frac{2}{1+4\mu}+\frac{5}{2(1+5\mu)}$  \\
\vspace{-3mm}
{} & {} & {} & {}\\
\hline%
\vspace{-3mm}
{} & {} & {} & {}\\
\vspace{-3mm}
 $5$ & \hspace{-2mm}$\lambda_5=\frac{2+12\mu}{1+7\mu} $ & \hspace{-3mm}
 $\rho_5\!=\!\frac{1+4\mu}{9+40\mu}\!\left(\!\frac{11+60\mu}{1+6\mu}\!-\!\frac{20(1+6\mu)}{1+7\mu} \!\right)$ \hspace{-3mm}&
 \hspace{-2mm} $E_5=\frac{5}{2(1+5\mu)}+\frac{3}{1+6\mu}$ \\
{} & {} & {}& {} \\
\hline%
\end{tabular}

\section{Mixed cases of deformed oscillators}\hspace{5mm}
It is possible to construct new models of deformed oscillators which
are quasi-Fibonacci extensions of a particular Fibonacci oscillator
of Section 2, by combining the latter with the $\mu$-oscillator
which plays the role of the basic quasi-Fibonacci "building block".
Clearly, the procedure brings in some additional deformation
parameters. On this way we naturally obtain new models with two
deformation parameters $\mu$ and $q$: say, the mixed $\mu\!-\!AC$
case, and also
the mixed $\mu\!-\!BM$ and $\mu\!-\!TD$ cases. However, since the
models of $q$-deformed oscillators are contained as particular cases
in the $\!p,q$-family, see sections 2.1-2.3, it is useful to start
with the three-parameter, mixed $(\mu;p,q)$-family of models.

\begin{center}
\vspace{3mm} {\it The three-parameter "mixed"\ family of
$(\mu;p,q)$-deformed oscillators} \vspace{1mm}
\end{center}
\hspace{5mm}This family of deformed oscillator models arises due to
combining the $p,\!q$-oscillator with the $\mu$-oscillator and is
given by the structure function
\begin{equation}
\varphi_{\mu,p,q}(n)=\frac{[n]_{p,q}}{1+\mu n} \ , \hspace{12mm}
[n]_{p,q}\equiv\frac{p^n-q^n}{p-q}. \label{62}
\end{equation}
Of course, the choice (\ref{62}) of "mixed" deformation is not
unique: for the "mixed" structure function one could also use, say,
$\psi_{\mu,p,q}(n)=\frac{[n]_{p,q}}{1+\mu[n]_{p,q}}$ or
$\chi_{\mu,p,q}(n)=[\frac{n}{1+\mu n}]_{p,q}$. Our choice however
(i) is simpler from the viewpoint of further use in applications
like that in \cite{mu-bose} and (ii) better correlates with the
ideas of \cite{Jan}.

So we expect that this $(\mu;p,q)$-oscillator, with the energy
spectrum
\[
E_n=\frac{1}{2}\Bigl(\varphi_{\mu,p,q}(n+1)+\varphi_{\mu,p,q}(n)\Bigr),
\] obeys the QF relation.

It is clear that the $(\mu;p,q)$-deformed family cannot satisfy pure
Fibonacci relation, because of its $\mu$-component as carrier of the
QF property; only if $\mu=0$ it reduces to the pure
$(p,q)$-oscillator for which the status of Fibonacci oscillator is
recovered.

 So, consider the $(\mu;p,q)$-oscillator as
quasi-Fibonacci one.
 The QF property of it
will be certified if the explicit expressions for $\lambda_n$ and
$\rho_n$ are found such that the QF relation (\ref{30}) holds true:
\[
\frac{[2]_{p,q}[n+1]_{p,q}-pq[n]_{p,q}}{1+\mu(n+2)}+
\frac{[n+1]_{p,q}}{1+\mu(n+1)}= \nonumber
\]
\begin{equation}
=\lambda_n\biggl(
\frac{[n+1]_{p,q}}{1+\mu(n+1)}+\frac{[n]_{p,q}}{1+\mu n}\biggr)+
\rho_n\biggl( \frac{[n]_{p,q}}{1+\mu n}+
\frac{[n-1]_{p,q}}{1+\mu(n-1)}\biggr)  \ .            \label{63}
\end{equation}
Note that in the LHS the identity:
   \begin{equation}
[n+2]_{p,q}=[2]_{p,q}[n+1]_{p,q}-pq[n]_{p,q}  \label{64}
\end{equation}
 has been used. Instead of (\ref{63}) we may {\it equivalently} exploit the
following two relations:
\[
\bigl(\lambda_n-1\bigr)\frac{[n+1]_{p,q}}{1+\mu(n+1)}=
\frac{[2]_{p,q}
[n+1]_{p,q}}{1+\mu(n+2)}-K\frac{[n+1]_{p,q}}{1+\mu(n+1)}\ ,
\]
\[
\bigl(\lambda_n+\rho_n\bigr)\frac{[n]_{p,q}}{1+\mu
n}+\rho_n\frac{[n-1]_{p,q}}{1+\mu(n-1)}=
\frac{-pq[n]_{p,q}}{1+\mu(n+2)}+K\frac{[n+1]_{p,q}}{1+\mu(n+1)}\ .
\]
Let us emphasize that the arbitrary function $K=K(n;\mu,p,q)$
involves both the variable $n$ and the three deformation parameters.
Like before, it is the unspecified function $K(n;\mu,p,q)$ which
guarantees the equivalence with (\ref{63}).
 From the pair of equations, using $[2]_{p,q}=p+q$ as implied by
(\ref{62}), the desired solution does follow, namely
\begin{equation}
\lambda_n\equiv
\lambda_n(\mu,p,q)=1-K+[2]_{p,q}\frac{1+\mu(n+1)}{1+\mu(n+2)}\
 , \label{65}
     \hspace{5mm}
\end{equation}
\[
\rho_n\equiv\rho_n(\mu,p,q)=\frac{1+\mu(n-1)}{[n]_{p,q}(1+\mu(n-1))+
[n-1]_{p,q}(1+\mu n)}\ \times
\]
\begin{equation}
\times \Biggl\{K\biggl([n]_{p,q}+\frac{[n+1]_{p,q}(1+\mu
n)}{1+\mu(n+1)}\biggr)-[n]_{p,q}\biggl(1+\frac{[2]_{p,q}(1+\mu(n+1))+qp(1+\mu
n)}{1+\mu(n+2)}\biggr)\Biggr\} \label{66}
\end{equation}
This constitutes our general result for the $(\mu;p,q)$-deformed
oscillators. From this, by fixing $K(n;\mu,p,q)$ we can generate for
$\lambda_n$ and $\rho_n$ various special expressions.

Let us consider some special cases of (\ref{65}), (\ref{66}), with
fewer deformation parameters.

1) If $p=q=1$ and $\mu\neq0$, it is easily checked that one recovers
the expressions (\ref{55}), (\ref{56}) in Sec. 3.3 for $\lambda(n)$
and $\rho(n)$ of the pure $\mu$-oscillator, the latter being the
carrier of QF property.

2) If we  put $\mu=0$ in (\ref{65}), (\ref{66}), we have:
\[\lim_{\mu\to 0} \lambda_n(\mu,p,q) = \lambda(p,q)=\lambda=1-K(n;p,q)+[2]_{p,q}\ ,
\]
\[\lim_{\mu\to 0} \rho_n(\mu,p,q) = \rho(p,q)=\rho=
\frac{K(n;p,q)([n+1]+[n])-[n](1+pq+[2]_{p,q})}{[n]_{p,q}+[n-1]_{p,q}}\
.
\]
In the latter expressions $\lambda$ and $\rho$ still depend on $n$.
However, at $K=1$ we recover $\lambda=[2]_{p,q}$ and $\rho=-pq$ of
the pure $(p,q)$-oscillator in Sec.2. The cancelation of the
$n$-dependence in $\lambda$, see $K(n;p,q)$, and in $\rho$,
follows by applying the identity (\ref{64}).

3) Let $p=q^{-1}$ and $q\rightarrow1$. The corresponding values of
the coefficients then read
\[
\lambda_n=1-K(n;1,1)+\frac{2(1+\mu(n+1))}{1+\mu(n+2)}\ ,
\]
\[
\rho_n=\frac{1+\mu(n-1)}{1+2(n-1)(1+\mu
n)}\!\left(\!K(n;1,1)\frac{2(n+1)(1+\mu n)\!-\!1}{1+\mu(n+1)}\!-\!
\frac{4n(1+\mu(n+1))}{1+\mu(n+2)}\right)
\]
and if moreover $K=1$, we have from the letter:
\[\lambda_n|_{K=1}\!=\!2\frac{1+\mu(n+1)}{1+\mu(n+2)},
\]
\[
\rho_n|_{K=1}\!=\!\frac{1+\mu(n\!-\!1)}{1+2(n\!-\!1)(1+\mu
n)}\left(\!\!1\!-
\!2n+\mu\!\left(\!\frac{4n}{1+\mu(n+2)}\!-\!\frac{n+1}{1+\mu(n+1)}\right)\right)\
.
\]

4) Putting $\mu=0$, $p=q=1$ in (\ref{65})-(\ref{66}),
 for $\lambda_n$ and $\rho_n$ yields
\[\lambda_n|_{\mu=0}\!=\!3-K, \hspace{5mm} \rho_n|_{\mu=0}\!=\!\frac{K(2n+1)-4n}{2n-1}\ .\]
see Remark 2 for more details.

As mentioned at the beginning of this section, since we have at our
disposal the explicit formulae (\ref{65}), (\ref{66}) for the
3-parameter $(\mu;p,q)$-family, we immediately get the corresponding
results for the three distinguished two-parameter cases. Let us
illustrate this (recall (8), (10), and $[n]_{TD}\equiv nq^{n-1}$ for
the AC, BM and TD cases correspondingly):

 \vspace{2mm} $\bullet$ let $p\to 1$.
\ \ \ In this case, $ \varphi_{\mu,AC}(n)=\frac{[n]_{AC}}{1+\mu n}$
\hspace{12mm} (mixed $\mu$-AC case);

\vspace{2mm} $\bullet$ put $p=q^{-1}.$ In this case
$\varphi_{\mu,BM}(n)=\frac{[n]_{BM}}{1+\mu n}$ \hspace{11mm} (mixed
$\mu$-BM case);

\vspace{2mm} $\bullet$ put $p=q$. \ \ \ In this case
$\varphi_{\mu,TD}(n)=\frac{[n]_{TD}}{1+\mu n}$ \hspace{13mm} (mixed
$\mu$-TD case).

\vspace{1mm} \noindent
 We end with the comment on the hybrid case of six
parameters: the  $(\mu;p,q,\alpha,\beta,l)$-case.

\vspace{2mm}
{\bf Remark 5.}
  All the above treatment carried out for the three-parameter
$(\mu;p,q)$-family of deformed oscillators can be easily extended to
the case of six-parameter family which combines the 5-parametric
case (\ref{17}) and the formula (\ref{27}) of the $\mu$-oscillator
case.
  Also, one can put $\l=\alpha$ in order to have
unambiguous definition of the structure function, consistent with
the relation $E_n=\frac12\bigr(\varphi(n)+\varphi(n+1)\bigr)$.

\subsection*{Conclusions and outlook}\hspace{5mm}
In this paper we dealt with two essentially different classes of
deformed oscillators: the Fibonacci class and the (much more rich)
quasi-Fibonacci class. As our first result, we have proven that
besides the well-studied 2-parameter family of $p,q$-deformed
oscillators known \cite{Ar_Fib} as Fibonacci oscillators, there
exists more general, with three additional parameters, the
$(p,q,\alpha,\beta,l)$-family of deformed oscillators which does
also belong to the Fibonacci class. What is rather unexpected, the
coefficients $\lambda$ and $\rho$ for the 5-parametric family of
deformed oscillators depend on $p$, $q$ (as in the $p,\!q$-deformed
case) and else only on one parameter $\alpha$, from the remaining
three parameters.

On the other hand, according to our second main result reflected in
Proposition 2, the $\mu$-oscillator \cite{Jan} is not in Fibonacci
class, but belongs to the class of QF oscillators whose basic
feature is that the sequence of values of energy levels obeys the QF
relation, being as well linear and three-term (or two-step) one, but
involving the \underline{non-constant} coefficients $\lambda_n$ and
$\rho_n$ which depend besides the deformation parameter $\mu$ also
on the number $n$ of the energy level $E_n$.

The peculiar feature of QF oscillators is that the
$n$-dependent coefficients $\lambda_n$ and $\rho_n$ are obtained
non-uniquely, as it was demonstrated with the three different
methods of solving the eq.(\ref{30}). While the first two ways lead
to partial solutions, the third method is general one due to the
(arbitrary) function $K(n;\mu)$ being involved.

We have shown that the $\mu$-oscillator is not the unique one
possessing the QF property.  Indeed, it can be utilized as a basic
ingredient for constructing other families of QF deformed
oscillators which have additional deformation parameter(s).
 An example of the three-parameter family of $(\mu;p,q)$-deformed
 oscillators obtainable through combining the $p,q$-oscillator with the
$\mu$-oscillator has been considered.
 From that, a particular two-parameter families
(the $\mu,q$-oscillator as the Arik-Coon $\&$ the $\mu$-type hybrid,
the BM $\&$ the $\mu$-type hybrid, and the TD $\&$ the $\mu$-type
hybrid) of deformed oscillators naturally follow.
  We believe our results provide the basis for 
  constructing in a regular manner numerous new models of multi-parameter
  deformed nonlinear oscillators belonging to different extensions of
  the Fibonacci class, in particular, those possessing the quasi-Fibonacci property.
  It is also worth to note that interesting  {\it classes of
  quasi-Fibonacci oscillators} have been explored in our recent paper \cite{GR-5}.
 The polynomially deformed non-Fibonacci oscillators treated therein are rather amusing
as they can be viewed in three different ways: 1) as quasi-Fibonacci
oscillators; 2) as the oscillators obeying inhomogeneous Fibonacci
relation; 3) as $k$-bonacci oscillators. Unlike the classes just
mentioned, the $\mu$-oscillator studied in the present paper along
with its direct extensions admit only the quasi-Fibonacci type (way)
of description.

  For each Fibonacci oscillator there exist diverse quasi-Fibonacci
  extensions, see e.g. the note below (\ref{62}) and Remark 5.
  This fact, all that is said in the preceding paragraph, and the results of Sec.3, Sec.4
  clearly demonstrate that the class of QF models based on the 
  novel concept of quasi-Fibonacci oscillators is really rich and
  worth of detailed study. Although exploring of physical applications of
  quasi-Fibonacci oscillators is at the very beginning, we may quote
  the recent work \cite{mu-bose} on the $\mu$-Bose gas model employing
  $\mu$-oscillators.
  Let us finally note that in view of existing deformed fermionic
  Fibonacci oscillators \cite{AAA} it is worth to study, along
  the lines of the present paper, possible models of fermionic
 quasi-Fibonacci oscillators, as yet another class.

\section*{Acknowledgement}

The authors are thankful to I.M. Burban for interesting discussions,
and to the referees for valuable remarks and suggestions which led
to improved presentation. This research was partially supported by
the Grant 29.1/028 of the State Foundation of Fundamental Research
of Ukraine and by the Special Program of the Division of Physics and
Astronomy of the NAS of Ukraine.


\begin{thebibliography}{0}

\bibitem{Cha-Ja} 
 Chakrabarti R and Jagannathan R 1991 {\it J. Phys. A: Math. Gen.}  {\bf 24} L711

\bibitem{Ar_Fib} 
 Arik M et al 1992 {\it Z. Phys. C} {\bf 55} 89--95

\bibitem{SCRM} 
 de Souza J, Curado E M F and Rego-Monteiro  M A 2006 {\it J. Phys. A: Math. Gen.} {\bf 39} 10415-10425

\bibitem{Schork} 
Schork M 2007 {\it J. Phys. A: Math. Theor.} {\bf 40}, 4207-4214

\bibitem{GR-5} 
 Gavrilik A M and Rebesh A P 2010 {\it J. Phys. A: Math. Theor.} {\bf 43} 095203 (15pp)

\bibitem{Chung} 
 Chung W-S, Chung K-S and Nam S-T 1993 {\it Phys. Let. A}  {\bf 183} 363-370

\bibitem{Borzov} 
 Borzov V V, Damaskinsky E V and Yegorov S B 1997 {\it Zap. Nauch. Semin. LOMI} {\bf 245} 80

\bibitem{Mizrahi} 
 Mizrahi S S, Camargo Lima J P and Dodonov V V 2004 {\it J. Phys. A: Math.
Gen.} {\bf 37} 3707

\bibitem{Burban}  
Burban I M 2007 {\it Phys. Let. A} {\bf 366} 308-314

\bibitem{Jan}    
 Jannussis A 1993 {\it J. Phys. A: Math. Gen.} {\bf 26} L233-L237

\bibitem{Wei}    
 Wei Shao-Wen, Ran Li, Yu-Xiao and Ji-Rong Ren {\it Quantization of
black hole entropy from quasinormal modes}, {\tt arxiv:0901.0587
[hep-th]}.

\bibitem{Man}   
Man'ko V I, et al. 1997 {\it Phys. Scripta} {\bf 55} 528

\bibitem{Mel}   
Meljanac S, Milekovi\'c M and Pallua S 1994 {\it Phys. Lett. B} {\bf
328} 55-59

\bibitem{Bona}  
Bonatsos D and Daskaloyannis C 1999 {\it Prog. Part. Nucl. Phys.}
{\bf 43} 537

\bibitem{AC}    
Arik M and Coon D D 1976 {\it J. Math. Phys.} {\bf 17} 524

\bibitem{Bied}  
 Biedenharn L C 1989 {\it J. Phys. A: Math. Gen.} {\bf 22} L873

\bibitem{Mcf}   
 Macfarlane A J 1989  {\it J. Phys. A: Math. Gen.} {\bf 22} 4581

\bibitem{AdG}   
Adamska L V and Gavrilik A M 2004 {\it J. Phys. A: Math. Gen.} {\bf
37} 4787

\bibitem{SIGMA} 
Gavrilik A M 2006 {\it SIGMA} {\bf 2} 12 (paper 074)  {\tt arxiv:
hep-ph/0512357}

\bibitem{Crn}   
 Crnugelj J, Martinis M and Mikuta-Martinis V 1994 {\it Phys. Rev. A} {\bf 50} 1785-1791

\bibitem{Algin} 
Algin A 2008 {\it J. Stat. Mech: Theor. Exp.} 10009

\bibitem{Algin2} 
Algin A 2009 {\it J. Stat. Mech: Theor. Exp.} 04007


\bibitem{Ben} 
Ben Geloun J, Govaerts J and Hounkonnou M N 2007 {\it Eur. Phys.
Lett.} {\bf 80} 30001 (6pp)

\bibitem{UFZh}   
Gavrilik  A M and Rebesh A P 2008 {\it Ukr. J. Phys.} {\bf 53} 586

\bibitem{Plethora} 
 Gavrilik A M and Rebesh A P 2008 {\it Mod. Phys. Lett A} {\bf 23} 921

\bibitem{Odaka}   
 Odaka K, Kishi T and Kamefuchi S 1991 {\it J. Phys. A: Math. Gen.} {\bf 24} L591

\bibitem{Jagan}  
 Chaturvedi  S, Srinivasan V and Jagannathan R 1993 {\it Mod. Phys. Lett. A} {\bf 8} 3727

\bibitem{GR-1}   
 Gavrilik  A M and Rebesh A P 2007 {\it Mod. Phys. Lett. A} {\bf 22} 949

\bibitem{AAW}    
Arik M, Atakishiyev N M and Wolf K B 1999 {\it J. Phys. A} {\bf 32}
L371.

\bibitem{d-Negro}  
 Luca Dal Negro and  Ning-Ning Feng 2007 {\it Optics Express} {\bf 15} No.22, 14396-14403

\bibitem{mu-bose}  
 Gavrilik A M and Rebesh A P, Intercepts of the momentum correlation functions in
$\mu$-Bose gas model and their asymptotics (submitted).

\bibitem{AAA}      
Algin A, Arik M and Arikan A S 2002 {\it Eur. Phys. J. C} 2002 {\bf
25} 487

\end{thebibliography}
\end{document}